\documentstyle[11pt]{article}
\newcommand{\blankline}{\vskip .3cm}
\newcommand{\f}{\begin{equation}}
\newcommand{\ff}{\end{equation}}

\begin{document}
\rightline{\Large CGPG-95/3-2 IASSNS-HEP-95/21}
\rightline{\Large gr-qc/9503027}
\vskip .3cm
\centerline{\LARGE  Experimental Signatures of Quantum Gravity }
\rm
\vskip.3cm
\centerline{Lee Smolin${}^\dagger$}
\vskip .3cm

 \centerline{\it   Center for Gravitational Physics and 
Geometry, Department of Physics${}^*$}
\centerline{\it Pennsylvania State University, 
University Park, Pa16802-6360, USA}
\centerline{and}
\centerline{\it   School of Natural Sciences, Institute for 
Advanced Study}
\centerline{\it Princeton, New Jersey, 08540, USA}

 \vfill
\centerline{revised version, November 1, 1995}
\vfill
\blankline
\centerline{ABSTRACT}
\noindent
\blankline
I review several different calculations, coming from string theory,
nonperturbative quantum gravity and analyses of black holes that
lead to predictions of phenomena that would uniquely be
signatures of quantum gravitational effects.
 These include: 1) deviations  from a thermal spectra for 
evaporating black holes,  2)  upper limits on the entropy and energy 
content of bounded regions,  3) suppression of ultra-high energy 
scattering amplitudes, consistent  with a modified uncertainty 
principle,  4)  physical volumes and areas have discrete 
spectra,  5),  violations of $CPT$ and universal violations of 
$CP$, 6) otherwise inexplicable conditions on the initial state of the 
universe or otherwise inexplicable correlations between cosmological 
and microscopic parameters.   Consideration of all of these together
suggests the possibility of connections between perturbative and 
nonperturbative approaches to quantum gravity.
\noindent
 \vfill
$\ \ {}^\dagger$  smolin@phys.psu.edu \ \ \  * permanent address
\eject

\section*{Introduction}

It is unfortunately true that we do not yet have a 
completely satisfactory quantum theory of
gravity.  At the same time, there has been a great deal of progress
in recent years in more than one approach to quantum gravity.  
One measure
of this progress is that we now know how to do 
several different kinds of calculations which describe
what might be called characteristic quantum gravitational 
phenomena.
These are conjectured phenomena that share the following 
features,  
 
1)  They seem to be follow necessarily from 
rather general assumptions about how to extend the basic 
principles of relativity, quantum theory and quantum field theory.

2)  They  would be impossible in a world described 
either by 
classical general relativity or by quantum field theory on a fixed 
spacetime background.

3)  In spite of the incompleteness of our present understanding
of quantum gravity, we are able to make specefic predictions 
concerning them which could be tested experimentally, subject
only to the technical feasibility of 
Planck scale experiments.

It is significant that 
these developments describe different kinds of phenomena, and
arise in the framework of  different approaches to
quantum gravity.  However, I do not think this should necessarily
imply that they could not all be simultaneously true.   While there
is as yet no complete theory of quantum gravity, it is possible
that the different approaces that to some partial extent succeed
are complementary rather than contradictory, in that they explore
different physical domains.
For example,
most calculations in string theory that bear on quantum gravity
are perturbative and describe small deviations from a
classical background, while most interesting
results in the loop representation approach to
quantum gravity are non-perturbative.  
Furthermore,  in each case the striking  results seem to rely 
more on general
principles, and less on specifics such as dynamics or matter content.  
In the case of string theory, predictions
about hyper-Planck scale 
physics\cite{he,attickwitten,string} 
follow rather directly when one
assumes that perturbative finiteness can be accomplished 
without
breaking Lorentz invariance\cite{lenny-lorentz,ks}.  In the 
case of the loop representation 
approach\cite{cr-review,aa-review,ls-review}, certain 
effects seem to follow rather directly
from the assumption that the representation and regularization
procedures are compatible with diffeomorphism 
invariance\cite{weaves,ls-review,volume}.  In
neither case are detailed dynamical assumptions involved.
Thus, it is
possible to imagine that at least some of the 
effects predicted by string theory are
characteristic of the perturbative domain of any consistent quantum
theory of gravity, while some of the effects found in the loop
representation approach are characteristic of any successful
approach to the non-perturbative domain. 

In addition, other
calculations, such as those showing violations in the 
thermal spectra of black holes\cite{ted-bh,unruh-bh,frankper,bekmuk},
seem  to rely only on rather
general assumptions
about the Planck scale, without making detailed assumptions
about the dynamics.

For this reason I would like to suggest that we try to see if we
can learn anything if we take the point of view that these 
different 
approaches are complementary,  rather than competing.
To this end, it may be interesting to list together some of the
different experimental predictions that have emerged out of
these different approaches, and to ask whether, taken together, they
might make a kind of a picture of physics at the Planck scale.
I thus describe 
in the following six sections
six different kinds of results which may be taken to describe
signatures of quantum gravitational physics.  

Unfortunately, due to limitations of  
space, I cannot attempt to give a complete survey
of all the experimental predictions that have come from
attempts at quantum gravity, nor do I give a complete list
of references in the cases I do describe.   The discussions
are abreviated, but details may be found in the cited
references.  Also, I apologise if 
more space is given to work with which 
I am more familiar; this is only
because my ignorance of some topics does not permit
me to discuss them in the same depth as the others.

\section{Corrections to the spectra of evaporating black holes}

In the semiclassical approximation, black holes have thermal
spectra\cite{hawking}.  As has been pointed out by many people, it is quite
likely that in a full quantum theory of gravity there must
be corrections to the thermal spectra of black holes.  These 
corrections are quite likely necessary to resolve the questions,
such as the information loss paradox,
raised by the semiclassical claculations; to this extent experimental
observations may in principle decide between different proposals
about how these puzzles 
are to be resolved\cite{bhpuzzle}.  Motivated primarily by these,
at
least three kinds
of corrections to the thermal Hawking spectra, 
coming from quantum gravity effects, have been investigated.  

The first kind 
follows from the hypothesis that information is not actually lost
in black hole evaporation.  If this is the case there must be 
corrections to the spectra coming from correlations between quanta
emitted from the black hole at different times\cite{correlations}.  

The second source of corrections to the Hawking radiation
comes from the peculiar fact that
the modes that dominate the Hawking radiation at times long
after the formation of the black hole are extraordinarily
blue shifted when they propagate near the horizon\cite{hawking}.  
Thus,
the derivation makes it seem as if effects at scales much shorter
than the Planck length are involved in the Hawking radiation.

On the other hand, all of the approaches to quantum gravity that
have been at least partially succesful require strong modifications
in the physics below the Planck length, which introduces some
truncation or diminuation in the number of degrees of freedom
at such scales\cite{garay}.   For example, one
such modification might be the presence of a
discrete spacetime structure at the
Planck scale.  Another might be a modification in the
energy-momentum dispersion relationship at hyperplanck
scales.  It is then very important to investigate whether such
changes from the naive free field behavior might lead to
modifications in the thermal spectra predicted by the semiclassical
theory.

This question was investigated from 
different points of view 
by Jacobson\cite{ted-bh} and Unruh\cite{unruh-bh}.  A key question
that arises in this work is whether there the existence of a discrete
spacetime structure can be consistent with Lorentz
invariance.  If not, there must in some circumstances
be prefered observers who see the violations from the
free field behavior to happen at a particular scale.
Jacobson hypotheses that 
the answer is yes, and imposes a cutoff in the frame of an 
observer freely falling into the
black hole.  The result is that there are small modifcations away
from a thermal spectra for evaporating black 
holes\cite{ted-bh}.  However,
even drastic changes in Planck scale physics do not lead to
a supression of the Hawking radiation for larger than Planck
scale black holes.
 
The work of Unruh was based on numberical studies 
of his model of sonic black holes, or dumb holes, in which horizons
appear in supersonic fluid flows\cite{unruh-bh}.  
He investigated whether 
modifications in the dispersion relation of sound waves at high
frequencies lead to modifications in the ``Hawking" radiation coming
from the hole.  He found that the radiation is still thermal.  However
his results do not rule out the possibility that there are small 
modifications in the spectra due to the short distance effects. 

These results suggest that the corrections to the spectra of
evaporating black holes are of the order of $m_{Pl} / M_{bh}$.
However there is a very interesting suggestion that this may be
too pessimistic, and that quantum gravity may actually induce
corrections in the predictions for the spectra of evaporating
black holes which are of order unity, no matter how large the
mass of the black hole is. In a very provocative paper,
Bekenstein and Mukhanov\cite{bekmuk} show that the simple
hypothesis that the area of the event horizon of the black hole
is quantized in discrete Planck scale units results in a rather
different spectra than that predicted by Hawking.  

The argument for this is quite simple.  Let us make the simplest
possible assumption for the quantization of area, which is that
the area of the horizon must take one of the values,
\f
A_{h}= n \alpha l_{Pl}^2
\ff
where $\alpha$ is a dimensionless constant of order one and $n$ is
an integer.  It follows that there is a discrete
spectrum of neutral non-rotating black holes
with masses,
\f
M_{bh}= m_{Pl} {1 \over 4} \sqrt{\alpha \over \pi} \sqrt{n}
\ff
It follows from this that an evaporating spherical black hole
may only radiate in integer multiples of a characteristic
frequency
\f
\omega_0 = {\alpha \over 16 \pi} {1 \over 2GM}
\ff
as long as it only makes transitions to other non-rotating black
holes.  There will, of course, be fine structure in the spectra of
black holes, due to the quantized angular momemtum.  The general
expression for a quantized Kerr black hole with quantized horizon
area and angular momentum is, using the formula of Christodolou and
Ruffini\cite{cr}
\f
{M^2 \over m_{Pl}^2 } =   {\alpha \over 16 \pi} n + 
{4 \pi \over \alpha }{l(l+1) \over n}
\ff
As long as the angular momentum is small compared to the irreducible
mass squared (in units $G=c=1$), the result will be a 
fine structure of the lines
coming from the black holes making transitions in both spin and area.
However, the resulting broadening of the of the lines will be
small compared to the spacing given by (3) as long as 
$J /M^2_{irr}$ is small.  Thus, except for the case of near-extremal
black holes, the broadening due to angular momentum cannot be
responsible for spreading out the lines sufficiently to recover the
continuous thermal spectra predicted by the semiclassical calculation.

The smallest frequency of the spectra (3) is 
near the peak of the thermal radiation.  This means that
if radiation could be seen from any evaporating black hole, there
would, according to the hypothesis of quantized area, be a stark
difference from Hawking's prediction of a continuous thermal
spectra.  

Bekenstein made the hypothesis that the area of black hole
horizons should be quantized some time ago\cite{bek-quant}.
More recently, the quantization of the areas of physically 
distinguished surfaces has been shown to be a
consequence of the loop representation approach
to non-perturbative quantum gravity, as I will describe in section 6
below.  Although the predictions of the spectra differs to 
some extent from 
equation (1), it is possible then that line emission of the sort
described by (3), rather than the continuous thermal spectra of
Hawking, must in fact be a general prediction of a nonperturbative
quantum theory of gravity.

Finally, one expects corrections 
to the thermal radiation in any quantum theory of gravity 
coming from the fact that the metric itself is quantized.
Kraus and Wilczek have been able to compute such corrections,
in the model in which the gravitational field is taken to be
spherically symmetric\cite{frankper}.  Because quantum gravity is expressed
in terms of constraints, the physical, gauge invariant perturbations
of a black hole involve necessarily functions of both the matter
fields and gravitational fields.  One cannot in a gauge invariant
way separate them from each other.  Kraus and Wilczek are
able to compute Hawking radiation in a model in which
gravity is coupled to matter fields, but spherical symmetry
is imposed.  They find that there is Hawking radiation in the
physical, coupled gravity/matter modes.  However, there are
corrections to the Hawking formula coming from the fact that
the metric is correctly included as part of the quantum modes.
This is a highly significant result, and should be quite independent
of which theory correctly describes Planck scale physics.

\section{Violations of $CPT$ and $CP$}

The $CPT$ theorem depends on assumptions about the geometry
of Minkowksi spacetime.  Furthermore, we do not expect  local
causality to hold in a quantum theory of gravity, because there is
no fixed background metric within which the fields propagate.  It
then may be expected that $CPT$ could be violated in a quantum
theory of gravity.

More specific reasons for expecting such a violation 
have come from two
directions.  First, Hawking proposed that $CPT$ must be violated
if information is lost in black hole evaporation\cite{hawking-cpt}.  
To discuss this work we must separate two claims made
by Hawking, the first, that $CPT$ is violated in the evaporation
of real black holes, and the second, that it is also violated to
some extent in all processes do to contributions from virtual
black holes.   While the first may imply the second, given certain
assumptions, it is also possible to imagine that in quantum gravity
structures such as black holes do not contribute to virtual
processes in the same way that field theory 
quanta do\footnote{One reason for this, recently
pointed out by Martinec\cite{martinec}, is that if there is a bounce
at Planck scales such that black hole singularities are
replaced by new, expanding regions of spacetime that
grow large (as in section 7, below) the quantum amplitude
for creation of a black hole is strongly 
suppressed.}.

That $CPT$ is violated universally due to virtual processes 
involving information lost in black holes is problematic
as it may, under certain assumptions, lead 
to large violations of energy\cite{bps}.  (But, for a criticism of this
view, see
a recent paper by unruh and wald\cite{uw-cpt}.)  
But these arguments do not
necessarily
bear against the possibility of $CPT$ violation in the
evaporation of real black holes.

 A different argument that quantum gravity may
violate both $CP$ and $CPT$ has been given by 
Chang and Soo\cite{changsoo-1,changsoo-2,chopin-cpt}.
Their results are based on the Ashtekar formulation of quantum
gravity\cite{aa-review}, treated in a 
path integral formulation.  While the
Ashtekar formulation is classically equivalent to general relativity,
they find that the two theories may differ quantum mechanically
due to the different ways they couple  the spacetime connection 
to chiral fermions.    In the Ashtekar formalism, the total action,
taken in the Jacobson-Samuels-Smolin form\cite{jss},
fails to be $CPT$ and $CP$ invariant when boundary terms 
are included.
The change of the action under 
$CP$ and $CPT$ is given by a chiral anomaly, coming
from the coupling of the left handed $SU(2)$ spin connection
to the chiral fermions.  In the presence of 
gravitational instantons, then
the full action is then not hermitian, and violations
of $CP$ and $CPT$ are expected.  

The possibility of $CP$ violation in quantum
gravity was also studied using the Hamiltonian formulation 
by Ashtekar, Balanchandran
and Jo\cite{abj}.  They showed that  quantum gravity may
naturally violate $CP$ through the appearance of $\theta$
vacua, with the left-handed spin connection playing the
role of the gauge field. We may note that as the effect comes
from the coupling of the spacetime connection to fermions,
any such gravitational violations of $CP$ will be universal.  This
universality might be used to distinguish quantum gravitational
$CP$ violating effects from other $CP$ violating effects.

\section{Limitations on the spectrum of fermions}

When there are an unequal number of left and right handed fermion
modes, as does appear to be the case in nature, the coupling
of the fermions to the spacetime connection need not
be  invariant under
a chiral, or partity, transformation that exchanges the left 
and right handed spacetime connections.  This is the
case in the Ashtekar formulation.
The result is that there
can be a global $SU(2)$ anomaly in the theory, coming from the
$SU(2)$ left handed spacetime connection.  This anomaly will
lead to contradictions if the path integral for quantum gravity
coupled to the fermions must receive contributions from manifolds
with arbitrary topology.   

This problem was studied by Chang and 
Soo \cite{changsoo-2}, who found that
if manifolds with arbitrary spacetime topology are to appear in the
Euclidean continuation of the path integral, there must be restrictions
on the fermion content of the theory.  Both the standard model
with $SU(3) \times SU(2) \times U(1)$ and the minimal $SU(5)$
grand unified model are ruled out.  The simplest grand unified
model that is allowed is $SO(10)$ with one $16$ dimensional
multplet of  Weyl fermions per generation.

Of course, it is not obvious that the correct theory of quantum 
gravity will have amplitudes that are representable by a
Euclidean path integral in which arbitrary topologies are
included.  One thing we learn from this analysis is that, if experiments
confirm that the correct description of matter is 
a grand unified theory which is inconsistent with these
conditions, any theory of quantum gravity that is representable
in terms of such a Euclidean path integral with a sum over topologies
and a chiral gravitational connection
is ruled out.  In this sense experiments in elementary particle
physics test hypotheses about quantum gravity.

\section{Scattering at transplankian energies}

String theory has provided us with the only consistent
semiclassical perturbation theory we have so far that can 
incorporate gravitation.  This does not necessarily mean that the
final theory of quantum gravity is a string theory, but it may mean
that in any consistent non-perturbative formulation of quantum
gravity, whether that be a string theory or not, the spectra of small 
oscillations around a semiclassical ground state must resemble that
of a string theory.  It is then extremely interesting that 
rather general 
arguments suggest that in string theory there is a cutoff scale
above which many fewer degrees of freedom are excited
than would be the case in a naive field theory.   These arguments
are based on 
several different analyses of 
physics at ultrahigh energies\cite{he,attickwitten,string}.
The same kinds of arguments have also been
used to suggest a universal modification of the uncertainty
principle, of the form,
\f
\Delta x > {\hbar \over \Delta p} + l_{Planck}^2 \Delta p
\ff
If such a relation holds than it is impossible to resolve any
structure on lengths shorter than the Planck scale.

A key issue which is probed in these experiments is the one I
raised above: whether the existence of a cutoff scale at Planck
energies, or a modification of spacetime structure, such that there
are many fewer degrees of freedom than in a conventional field
theory below the Planck scale, can be consistent with Lorentz
invariance.  It seems that in string theory this is achieved
in a very interesting way, which leads to the
modified uncertainty relation (1).  Very interesting discussions
of this issue are found in the papers of Susskind and 
collaborators\cite{lenny-lorentz}, 
who show why in a string theory Lorentz invariance can be 
compatible with the existence of a finite cutoff scale.  It
may, indeed, be exactly because of this that string theory succeeds
in giving a consistent perturbative description of quantum 
gravitational interactions.

\section{Limitations on the information and energy
content of finite regions of space}

Another characteristic limitation on the numbers of degrees of
freedom that can be excited in a finite region comes from the
existence of black holes.  If more energy than $\sqrt{A}$ 
(in Planck units) is
put in a region surrounded by a boundary of area $A$, we may
expect that a black hole will form.  This has several implications
that would lead to characteristic tests of quantum gravity.  First,
we may conjecture that there is an upper bound for the energy
that can be contained in any region, which is given by $\sqrt{A}$.
We may note that this may be shown for the spherically
symmetric case, assuming only the positive energy
conditions\cite{unpublished}, and it may be conjectured 
to hold in general. 

We may also conjecture that as a result of this upper limit
on the energy, the Hilbert space for quantum theory for any finite
region will be finite dimensional, because there is then present
both a low and high frequency cutoff.  Related to this conjecture
is the Beckenstein conjecture that the information that can be
contained within any finite region is bounded by its area, in Planck
units\cite{bek}.  
We may note that if this is the case then, as proposed by
't Hooft\cite{gerard-holographic} 
and Susskind\cite{lenny-lorentz}, the physical 
state describing any bounded
region of space should be describable in terms of a finite field
theory on the surface of that region.  They call this 
the {\it holographic hypothesis.}

Completely independent arguments that in quantum gravity the 
quantum state describing a region is actually a quantum state of
a field theory on its boundary have been given by Crane, in the
context of an analysis of the role of diffeomorphism invariance
on the interpretation of any quantum theory of 
gravity\cite{louis-surfaces}.   
These arguments have been strengthened
recently by progress in the construction of four dimensional
topological quantum field 
theories\cite{louis-4dtqft}, based on this picture.
Additionally, evidence that some, and perhaps all of the information
contained in a quantum state of 
in general relativity in a region bounded by a finite boundary
has been found by the author, at least in the Euclidean case and only for
a particular choice of boundary conditions\cite{cs-surfaces}.

Thus, completely independent lines of argument lead to the
conclusion that in quantum gravity the number of degrees of
freedom which may be observed
inside a region bounded by a surface of finite area must
be finite and bounded.
One important effect of this hypothesis will be on the 
thermodynamics of the quantum gravitational field, coming
from bounds on the entropy and energy of the contents
of such bounded regions.  These 
implies that the very high temperature thermodynamic 
behavoir of any finite region
will be very different than would be expected in conventional
quantum field theory.  Thus, we may expect that, were it possible
to heat an oven up to temperatures on the order of the square root
of its area, (in Planck units)  quite spectacular results would be
achieved\footnote{One caution about such experiments that must
be mentioned is the impossibility of constructing an oven that 
would contain gravitational radiation\cite{nouvcatastrophe}.}.

Finally, we may note that, as far as we know, the region of the
univese within our present horizon was at one time contained in
a region bounded by a very small area.  Perhaps this is the real
explanation for the horizon and flatness problem, or, in other
words, for Penrose's 
{\it Weyl curvature hypothesis}\cite{roger-weyl} according
to which the big bang singularity is characterized by very
low entropy.  For if Beckenstein's bound is true, then the
state space that describes the possible initial conditions of a 
region of the universe that is initially only the Planck
scale is very small. 

\section{Discreteness of area and volume}

A major theme of work in quantum gravity from the beginning has 
been the hypothesis that the combination of the principles of general 
relativity and quantum theory leads to a discreteness of the 
geometry of space or spacetime at Planck scales.   The many 
different directions from which this conclusion has been reached are 
described in a recent article, which is recommended to the 
reader\cite{garay}.  Here 
I would like to describe one set of predictions
concerning such a discrete structure that have recently been 
obtained\cite{volume,ls-review}, which is based on the loop representation
formulation of quantum gravity\cite{lp1,gambini}.

What is found here is that given certain general assumptions about
the quantum theory of gravity, it is not possible to define operators
that measure local quantities in the theory, such as the
components of the
metric or the curvature tensor at a point\cite{ls-review}.  
Instead, physical meaning
can only be given to certain non-local operators.  Furthermore, in
at least two cases, these non-local observables, which have 
continuous spectra in classical general relativity turn out to have
discrete spectra in the quantum theory\cite{weaves,volume}.  
Moreover, given only 
rather general kinematical assumptions, these spectra may be
computed.

The two cases for which this has been so far worked out
are the area of any physical two dimensional 
surface and the volume of any 
physical three dimensional 
region.  Here what I mean by a physical surface or region
is one defined by the values of some physical fields.  The particular
three dimentional surface in spacetime within which these 
surfaces and regions are defined may be picked out by some 
physical field such as a scalar 
field\cite{ls-brill}, after which a two surface
may be picked out by the value of a 
second field\cite{mc,matterrefl}.  
The physically
meaningful areas is then the area of that two dimensional surface
or the volume of that three dimensional region.

Thus, the observables we are discussing are composite operators,
that involve measuring simulatanously matter and
gravitational fields.  As such they are not defined naively and
must be defined through a regularization proceedure.  

As a result of the necessity of regularization, a key
requirement of the quantum theory of gravity emerges as
greatly restricting the framework of the theory. This 
is {\it diffeomorphism invariance.}  This appears
to place several limitations
on the quantum field theory which is being used.  

First of all, diffeomorphism invariance places
limitations on the representation of the observable algebra on
which the quantum theory is based\cite{ls-review,math}.  The 
representation must
be one that allows us to construct  nonanomolous generators of
the three dimensional spacial diffeomorphism group.  At present,
only one kind of represntation is known to have this 
property, which is the loop representation, with a discrete norm
of the type described in \cite{rayner}, and developed 
rigorously by\cite{math}.
Although there is not yet a theorem that these are the only
representations of the observable algebra of quantum gravity that
will allow imposition of the generators of diffeomorphisms, it is
quite possible that this is the case.  In particular, Fock
spaces based on a fixed spacetime metric do not have
this property.  For this reason, we work with
these representations.

The second limitation is that the composite operators that represent
observables such as we have been discussing must be constructed
through a regularization procedure that does not violate the
diffeomorphism invariance of the representation used to 
construct the quantum field theory.   As we are describing
only kinematical observables that do not involve the dynamics,
what needs to be done at this stage is something analogous to
the normal ordering of conventional Fock space quantum field 
theory.  We may recall that, in the case of Fock space, normal
ordering reveals the physical content of the theory, which in
the case of free field theory is the spectra of quanta.  In the
case of quantum gravity, we do not have a systematic account
of possible regularization procedures, but so far only one
kind of procedure has been found which does not violate
diffeomorphism invariance\cite{ls-review}.  This leads to definite
predictions, which is that the areas of physical surfaces
and the volumes of discrete regions come in certain discrete
spectra\cite{weaves,volume}.  To 
understand these, it is necessary to know that the
spectra of states which arises from the diagonalization of these
operators are in one to one correspondence with certain 
graphs, which are called spin networks\cite{volume,network}.  
These are graphs
in which the edges are labled by integers, that label $SU(2)$
representations.  For the case of the area of a surface, the spectra is
then\cite{volume}
\f
A =  \ l_{P}^2\ \sum_{l} \sqrt{j_l (j_l +1)}
\ff
where the sum is over the intersections of the edges of the graph
with the surface and $j_l$ is the spin of the $l$'th edge when it intersects
the surface.

In the case of the volume, the spin network states still provide
a basis with diagonalizes the volume operators.  In a few cases
the eigenvalues of the volume have been computed explicitly.
It turns out that the volume contributed by any trivalent
vertex is zero\cite{renata}\footnote{This corrects a sign error
in the original calculation \cite{volume}.}.  The eigenvalues
associated with some higher valence vertices have been 
computed\cite{renata,qdeformed}.
Calculations underway now, but presently incomplete, 
are expected to lead to a general
expression form the spectrum of the volume operator.   
 
We may note that
as we are describing
observables that measure the properties of the geometry of
a given three surface, no dynamics is involved. Thus no assumption
is made about the field equations or the constraints that describe
the evolution of the fields in time.  Thus, these predictions must hold
in any field theory describing gravity that may be expressed in
terms of frame field and connection variables.  In particular,
these spectra stand as much as predictions of supergravity,
higher derivative quantum theories or dilaton theories as they
are of conventional general relativity, and they are independent of
what matter fields are coupled to the theory.

Several recent results also attest to the robustness of the predictions
of the discreteness of area and volume observables in non-perturbative
quantum gravity.  Ashtekar and collaborators have found similar,
if not identical, results for the quantization of the area in the context
of a mathematically rigorous formulation of quantum general relativity
based on the connection representation\cite{gangof5}.  Renata Loll
has found that the volume is quantized also in a lattice formulation
of quantum gravity\cite{renata}.  
Finally, it has been found recently that the
whole formulation of quantum gravity in terms of 
spin networks\cite{network} extends naturally to a $q$ deformed
representation\cite{qdeformed} in which the deformation 
parameter is $q=e^{\imath \pi / (k+2)}$,
where $k=6\pi /G^2 \Lambda$. 

\section{Predictions about the parameters of the standard model of 
particle physics}

One of the things that we would like a quantum theory of 
gravity to do for us it to tell us what happens to the singularities
that are predicted by classical general relativity.  There are, roughly
speaking, three possibilities as to what happens when quantum
physics is taken into account at both black holes and cosmological
singularities.  1)  The singularity is removed and time continues
without bound.  2)  The singularity persists, even in the quantum
theory.  3)  Something new and unexpected happens to time, for
example time simply becomes ill defined in such regions, as in
the proposal of Barbour\cite{julian-time}.

It is interesting to note that in the first case, we can make hypothesis
which can lead to predictions that are testable by means of
astrophysical observations and theory.  This is because of the 
possibility that in our past there lies a succession of events in 
which a region of the universe collapsed almost to singularity
and then expanded again.

Unfortunately, it is still not the case that we have a theory that can
tell us what happens to singularities in our real, four dimensional
spacetime.  But, two simple hypotheses about the fate of singularities
may be stated that do have testable 
consequences\cite{evolution1,evolution2}.  These are

1)  {\it The bounce hypothesis:} Quantum effects cause collapsing
matter to begin again to reexpand whenever the density of matter
approaches the Planck density.

2)  {\it The mutation hypothesis:} Whenever a region of 
spacetime reaches a density or temperature near the Planck
scale, the dimensionless parameters that characterize low
energy physics change by small random amounts.  The new 
values hold for the future of that region.

To see how these leads to observable 
consequences, we may note that
the first hypothesis means that to the future of every
surface that, according to classical general relativity, would be a 
black hole singularity, there develops a new expanding region of
the cosmos.  This region is protected from view by observers in
the region where the black hole formed by the event horizon of
the black hole, for time scales less than the black hole evaporation
time. (We may note that for astrophysical black holes this is
enormously greater than the age of the universe.)  Conversely,
our expanding universe might, according to this hypothesis, be
one such region, expanding to the future of another region in which
a black hole formed.  Thus, the universe consists of many regions,
each separated from the other by the event horizons of the black
holes which lead from one to the other.

We may note that while such a scenario has been conjectured for
some time, a recent calculation that suggests that it is
actually a consequence of string theory\cite{martinec}.  Other
proposals have been made concerning the short distance behavior
of quantum gravity which also results in the bounce 
hypothesis\cite{brandenberger1}.

The idea that the laws of physics might change at such bounces is
not a new one, in the context of bounces of the cosmological 
singularity it has been called by John Wheeler ``the reprocessing
of the universe."  We may note that in the black hole
case, some form of cosmic censorship is required if the futures of
the almost singular region where the couplings change are not
to overlap.  We might note that this hypothesis is also consistent
with the present state of knowledge of string theory, which is
that we have a very large number of apparently equally
consistent perturbative string theories, apparently describing
small perturbations around different vacuum states, characterized
by different dimensions and low energy matter content.  The new
suggestion is then that all of these may be
realized in nature and that the principle that realizes which
of these is realized in our universe is statistical and reflects
contingent factors about the past of the interior of our
horizon.

What is interesting is that, as I shall now describe, if we 
restrict how the laws of physics can change at the bounces to 
small variations in the parameters of
the standard model of particle physics (expressed in terms of
dimensionless ratios) we find there are testable consequences.

These consequences follow because we can now make statistical
predictions about the regions\cite{evolution1,evolution2}.  
This is because it follows from these
assumptions that after a large but finite number
of iterations almost every region has a 
particular property, which is that its parameters are near to those
that extermize a certain quantity, which is the average
number of black
holes produced by a region with those values of the parameters.
This is testable because it is natural to assume that our
region of the universe is typical, in which case any property held
by a typical region must be a property of our observable universe.

To test this we have to combine astrophysical observation and 
theory, and ask whether increases or decreases of the paramters
of the standard model lead in almost every case to a decrease
in the number of black holes produced.  This hypothesis has been
examined, and a significant amount of evidence found in its favor.
For details the reader is referred to 
\cite{evolution1,evolution2,book}.  A partial list of
those changes that decrease the number of black holes formed is
1)  increasing the proton-neutron mass difference, 2) decreasing
$\alpha_{QCD}$, 3) increasing $\alpha$, 3) increasing $m_{electron}$,
4) increasing $m_{\nu_{e}}$, 5) increasing or decreasing
$G_{Fermi}$ 6) making $m_{proton} > m_{neutron}$, 7)  increasing
$\Lambda_{cosmological}$,  

A very interesting test is being examined at the present time,
in which the parameter varied is the mass of the strange quark.
According to a recently proposed theory of neutron star matter of Bethe and
Brown\cite{bethebrown}, 
an increase in the kaon mass above its present
value would increase the upper mass limit for neutron stars,
which would greatly decrease the number of black holes produced.
We are investigating at present whether decreases in the kaon
mass would not significantly increase the number of black holes.  If
this is the case it would stand as a strong confirmation of the 
hypothesis.

However, whether or not this scenario turns out to be true, the fact
remains that because the universe itself passes through Planck scale 
regions, hypotheses about quantum gravity can have cosmological
and astrophysical implications.  What I have described here is only
one way that natural assumptions about what happens at the Planck
scale might lead to testable predictions about astronomical observations.

\section{Conclusions}

The list of experimental consequences of theories of quantum gravity 
I have given here is incomplete, but even so I hope that
it makes the point.  We know quite a lot about what we might 
reasonably expect the characteristic experimental signatures of
quantum gravity to be.  Experimental observation of any of the
following phenomena would likely be useful as tests of a quantum
theory of gravity:  1)  deviations from a thermal spectra for
evaporating black holes,  2)  upper limits on the entropy and
energy content of bounded regions  3)  suppression of 
ultra-high energy scattering amplitudes, consistent with the
modified uncertainty principle (1)  4)  the discovery that
observables that measure aspects of the spacetime
geometry, such as physical volumes or the areas, 
have discrete spectra  5)  violations
of $CPT$ or universal violations of $CP$.  6) 0therwise inexplicable
conditions on the initial state of the universe or otherwise
inexplicable
correlations between cosmological and microscopic properties, such
as discussed in the previous section.

Further, while these predictions come from different theoretical
programs, in each case the predicted phenomena are rather robust,
and come from the most general assumptions about quantum gravity, and 
not from detailed assumptions about the form of
the dynamics or the matter content of the theory.  Nor do they
for the most part
depend on the very difficult foundational problems associated
with quantum cosmology.    Thus, I would venture to
make the optimistic statement that the
present situation in quantum gravity is perhaps analogous
to the quantum theory of about 1918, when physicists 
knew of a number of different characteristic 
quantum effects, in
atomic physics, thermal radiation and low temperature 
physics, without having yet a complete theory.

How are we then to go ahead and construct that complete theory?
The strongest thing that emerges from this list
of quantum gravitational predictions is that they all indicate in one
way or another that the beast we are after cannot be a conventional
quantum field theory, if by that we mean that there are an infinite
number of degrees of freedom within every physical volume.  
What is striking is also the way in which the results of these
calculations are often simpler than the machinery one has to employ
to derive them from the formalisms we have.  
Certainly
some of the steps of these calculations, 
such as the way diffeomorphism invariant states and operators
are built out of unphysical and infinite dimensional structures, or the
way that calculations in string theory are so far limited to expansions
around vacuum states associated with classical geometries,
reflects more our present
stage of ignorance than they actually mirror anything 
in nature.  But more than this, classical field theory, with its
infinite degrees of freedom, and enormous redundency of
degrees of freedom due to diffeomorphism and gauge invariance,
must be an approximation to reality
and not a starting point for the construction of the right theory.
Instead, in different ways the results of string theory, topological
field theory and nonperturbative quantum gravity suggest that the
right mathematics to construct quantum gravity is some combination
of algebra, representation theory, combinatorics and category theory.

At some point we may have to take a leap
and attempt to build the theory of quantum gravity up from some
simple structure we posit as a first
principle.   
What general conclusions can we draw about the Planck scale
that may then be clues for the construction of such a theory?
I would venture the following remarks:

1)  The physical fields must create non-local structures, most 
likely one dimensional.  This is the conclusion of both string
theory and non-perturbative quantum gravity.  Susskind's
analysis explains to us why this is necesssary if a
perturbative is to have a finite cutoff and at the same time
not break lorentz invariance\cite{lenny-lorentz}.  On the other hand,
non-perturbative quantum gravity explains why this is necessary
if we are
to have a quantum field theory based on a connection that does not
break diffeomorphism invariance.   This is an important example
of the way in which string theory tells us how  a quantum theory
of gravity must work in the perturbative domain, while the 
loop representation approach tells us how it must work 
non-perturbatively.  
There is something important yet to be understood in the fact that
both points of view lead to the conclusion that the physical 
excitations of the quantum gravitational field must be one dimensional.

2)  There are not an infinite number of degrees of freedom in any
volume.  There are instead a finite number of degrees of freedom in
every bounded region. 
Again, it is striking that
string theory reaches this conclusion from requiring the consistency
of the perturbative description, while the loop representation
reaches the same conclusion from the requirement of the consistency
of the non-perturbative description.  While the resulting pictures
are apparently quite different, it seems quite possible there is a
connection between them.   Moreover, several arguments discussed above
suggest that there are actually only a fixed number of 
degrees of freedom per unit area of the boundary of any region 
(expressed in Planck units.) 

3)  In both string theory and the loop representation,  constructed
as in \cite{ls-review,volume},
physical quantities are ``automatically" ultraviolet, 
finite but there are dangers of infrared divergences.  In
both cases the finiteness comes from the existence of a finite
number of degrees of freedom below the Planck scale.  The danger of
infrared divergences seems to have a rather different origin
in string theory and non-perturbative quantum gravity.  However, given the
relationship between the infrared divergences in bosonic string
theory and the danger of forming ``spikes" in random surface
theory they may in the end be related.  In both of these 
cases the problem is that finite Planck scale
quantum geometries  do not normally correspond to slowly
varying classical geometries.  To do so special conditions must
be met.

We may note that a similar situation was found in the Monte Carlo 
approaches to four dimensional quantum gravity within the
framework of dynamical triangulations\cite{4dtriangles}.  
For most values of the
gravitational and cosmological constants the theory does not seem
to have a critical behavoir that would result in correlation
functions behaving as if they live in a four dimensional classical
geometry.  To achieve this the theory must be tuned to a critical
point.  The fact that this problem is found also in  the other
non-perturbative approaches suggests that it is generic.  It is
possible that the solution is also generic.  This would imply that
in a non-perturbative quantum theory of gravity the classical limit, 
when it exists,  must be a critical phenomena\cite{cosmocritical}.

At the perturbative level, string theory finds a different solution
to the problem of the infrared divergences, which is by the
 addition of supersymmetry, which removes the tachyonic
divergences.   This suggests that it could be very important to
explore the implications of supersymmetry in the non-perturbative,
loop representation, approach\footnote{There is an extension
to supergravity\cite{ted-sg} which deserves more attention.}.  
There are, it might be added, 
important results from supersymmetric Yang-Mills theory that
show how supersymmetry can control the infrared behavior of
strongly coupled theories\cite{seibergwitten}.

But, beyond this, it is striking that in all the 
non-perturbative formulations 
of quantum gravity, the problem of the classical limit seems to be a 
problem of a critical phenomena.  The theory must find or be tuned
to a critical point for there to be a classical world at all.  In this sense
the existence of classical spacetime arises from a limit which has
much in common with the thermodynamic limit.   Perhaps in this circumstance
is to be found the connection between spacetime and thermodynamic
that is suggested by the results on black hole evaporation.

Finally, in closing it must be said that, 
unlike the case of atomic physics in the 1920's, 
the major impediment to progress in quantum gravity remains the
difficulty of doing experiments to check predictions like those
described here.  
But it can
no longer be said that theorists have been unable to make
experimental predictions about quantum gravity.  Perhaps it is
beginning to be time to wonder whether there might not be 
unforseen ways to test for the presence of these characteristic
quantum gravitational phenomena.

\section*{ACKNOWLEDGEMENTS}

I am grateful to Abhay Ashtekar,  Mauro Carfora, Jane Charlton,
Louis Crane,  
Ted Jacobson, Martin Rees, Carlo Rovelli, Chopin Soo,
Leonard Susskind and Gerard 'tHooft 
for discussion on the subject of this paper.  The 
opportunity to be a visitor at the Institute for Advanced Study has
been very helpful in formulating some of the suggestions made here.
This research was supported by the NSF under grant number 
PHY-9396246 and research funds provided by Penn State 
University.

\end{document}